\documentclass[]{spie}  

\usepackage{amsmath,amsfonts,amssymb}
\usepackage{graphicx}
\usepackage[colorlinks=true, allcolors=blue]{hyperref}
\usepackage{amsmath,amsfonts,amssymb}
\usepackage{graphicx}
\usepackage{setspace}
\usepackage{tocloft}
\usepackage{bm}

\title{Graph Embedding Using Infomax for ASD Classification and Brain Functional Difference Detection}

\author[a]{Xiaoxiao Li}
\author[a]{Nicha C. Dvornek}
\author[a]{Juntang Zhuang}
\author[a]{Pamela Ventola}
\author[a]{James Duncan}
\affil[a]{Yale University, New Haven, USA}

\begin{document} 
\maketitle

\begin{abstract}
    Significant progress has been made using fMRI to characterize the brain changes that occur in ASD, a complex neuro-developmental disorder. However, due to the high dimensionality and low signal-to-noise ratio of fMRI, embedding informative and robust brain regional fMRI representations for both graph-level classification and region-level functional difference detection tasks between ASD and healthy control (HC) groups is difficult. Here, we model the whole brain fMRI as a graph, which preserves geometrical and temporal information and use a Graph Neural Network (GNN) to learn from the graph-structured fMRI data. We investigate the potential of including mutual information (MI) loss (Infomax), which is an unsupervised term encouraging large MI of each nodal representation and its corresponding graph-level summarized representation to learn a better graph embedding. Specifically, this work developed a pipeline including a GNN encoder, a classifier and a discriminator, which forces the encoded nodal representations to both benefit classification and reveal the common nodal patterns in a graph. We simultaneously optimize graph-level classification loss and Infomax. We demonstrated that Infomax graph embedding improves classification performance as a regularization term. Furthermore, we found separable nodal representations of ASD and HC groups in prefrontal cortex, cingulate cortex, visual regions, and other social, emotional and execution related brain regions. In contrast with GNN with classification loss only, the proposed pipeline can facilitate training more robust ASD classification models. Moreover, the separable nodal representations can detect the functional differences between the two groups and contribute to revealing new ASD biomarkers.
\end{abstract}
\footnote{preprint,under review of SPIE Biomedical Imaging}


\section{INTRODUCTION}
 Autism spectrum disorder (ASD) affects the structure and function of the brain. Functional magnetic resonance imaging (fMRI) produces 4D spatial-temporal data describing functional activation but with very low signal-noise ratio (SNR). It can be used to
characterize neural pathways and brain changes that
occur in ASD. However, due to high dimension and low SNR, it is difficult to analyze fMRI.  Here, we address the problem of embedding good fMRI representations for identifying ASD and detecting brain functional differences between ASD and healthy control (HC). To utilize the spatial-temporal information of fMRI, we represent the whole brain fMRI as a graph, where each brain region (ROI) is a node, the underlying connection can be calculated by fMRI correlation matrix and node features can be predetermined, hence preserving both geometrical and temporal information. The Graph Neural Network (GNN), a deep learning architecture to analyze graph structured data, has been used in ASD classification \cite{li2019graph}. In addition to improving ASD classification, one core objective of our work is to discover useful representations to detect brain regional differences between ASD vs HC. The simple idea explored here is to train a representation-learning function related to the end-goal task, which maximizes the mutual information (MI) between nodal representation and graph-level representation and minimizes the loss of the end-goal task. MI is notoriously difficult to compute, particularly in continuous and high dimensional settings. Fortunately, the recently proposed MINE \cite{belghazi2018mine} enables effective computation of MI between high dimensional input/output pairs of a deep neural network, by training a statistics network as a classifier of samples coming from the joint distribution of two random variables and the product of their marginals. During training of a GNN, we simultaneously optimize the classification loss and Infomax loss \cite{velivckovic2018deep}, which maximizes the MI between local/global representation. In this way, we tune the suitability of learned representations for classification and detecting group-level regional functional differences. Results show the improvement of the classification task and reveal the functional differences between ASD and HC from the separable embedded brain regions encoded by the GNN.
\section{METHODOLOGY}

\subsection{Data Definition and Notations}
\label{define}
Suppose each brain is parcellated into $N$ ROIs based on its T1 structural MRI. We define an undirected graph on the brain regions $\bm{G} = (\bm{V},\bm{A})$, where $\bm{V} = ( \vec{v}_1,\vec{v}_2, \dots, \vec{v}_N)^T \in \mathbb{R}^{N\times D}$ and $\bm{A}= [a_{ij}]\in \mathbb{R}^{N \times N}$, and $D$ is the attribute dimensions of nodes. For node attributes, we concatenate handcrafted features: degree of connectivity, General Linear Model (GLM) coefficients, mean, and standard deviation of task-fMRI, and ROI center coordinates. $\bm{A}$ is calculated by the correlation of the mean fMRI time series in each ROI. Graph convolutional kernel (Section \ref{encode}) will encode the input graph to a feature map $\bm{H}=(\vec{h}_1, \vec{h}_2,\dots,\vec{h}_N)^T\in \mathbb{R}^{N\times F} $, that reflects useful structure locally. Next, we summarize the node representation into a global feature $\vec{s}$ by pooling and reading out (Section \ref{pool}). Given a $\bm{G}$, we will generate a negative graph $\bm{G'}$, whose embedded node representation is $\bm{H'}$. The corresponding positive pair $(\vec{h}_i, \vec{s})$ and negative pair $(\vec{h}'_i, \vec{s})$ will be encouraged to be separated by a discriminator $\mathcal{D}$ (Section \ref{discriminator}).

   \begin{figure} [ht]
   \begin{center}
   \begin{tabular}{c} 
   \includegraphics[height=4.5cm]{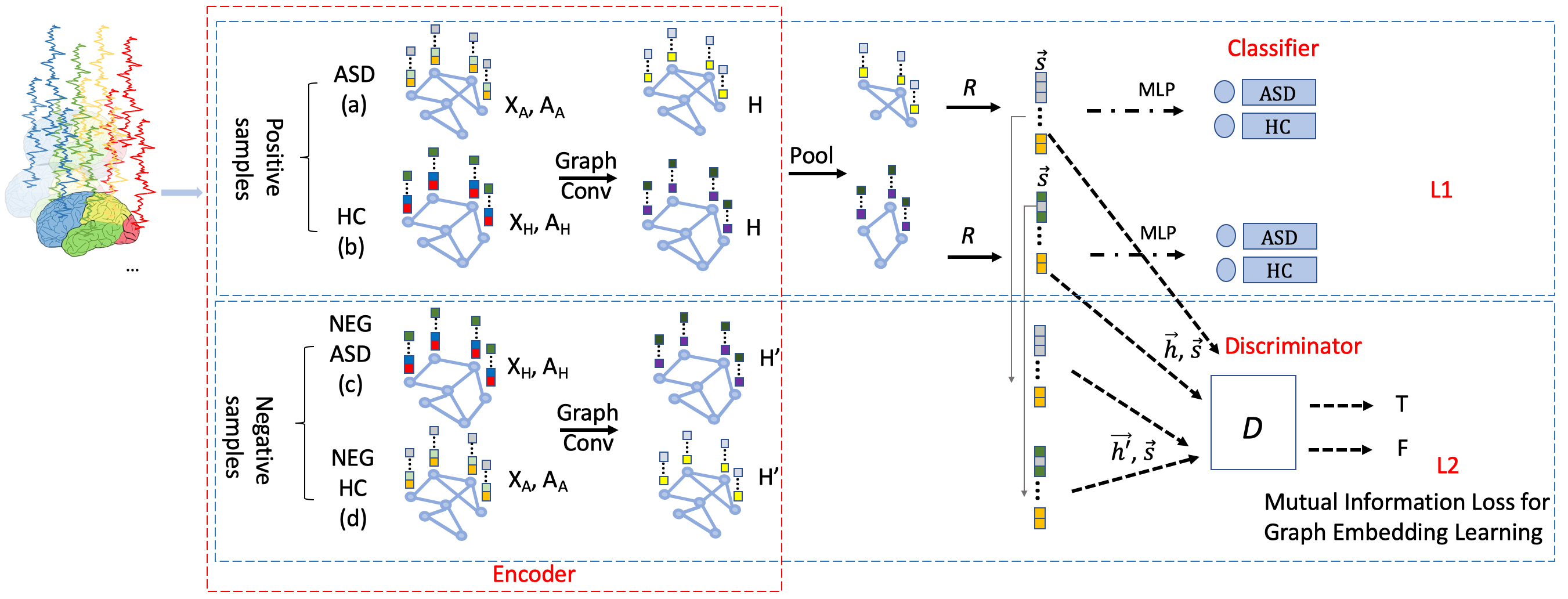}
   \end{tabular}
   \end{center}
   \caption[example] 
   { \label{flowchart} 
The flowchart of our proposed ASD classification and graph embedding architecture. The top row of the flowchart is a Graph Neural Network architecture to classify ASD and HC. The bottom row is a graph infomax pipeline to encourage better graph embedding. 
Here, (a) and (b) are positive samples; (c) and (d) are negative samples. (a)(c) (or (b)(d)) is a paired graph. The inputs of discriminator \textit{D} is the summary vector generated from positive samples, paired with node embedded representation ($\bm{H}$ or $\bm{H'}$). ($\vec{h}_i,\vec{s}$) pair will have True (T) output from \textit{D}, whereas ($\vec{h}'_i,\vec{s}$) will have False (F) output. The encoder, classifier and discriminator are trained simultaneously.}
   \end{figure} 
\subsection{Encoder: Graph Convolutional Layer}
\label{encode}
Our encoder $\mathcal{E}$ node embedding network is a $L$-layer supervised GraphSAGE architecture \cite{hamilton2017inductive}, which learns the embedding function $\mathcal{E}: \bm{X}\in \mathbb{R}^{N \times D}  \rightarrow$ $\bm{H} \in \mathbb{R}^{N \times F}$ mapping input nodes $\bm{X}$ to output $\bm{H}$. The embedding function is based on the \textit{mean-pooling} (MP) propagation rule $MP(\bm{X,A}) = \bm{\hat{D}}^{-1}\bm{\hat{A}}\bm{X}\bm{\Theta}$ as used in Hamilton et al.\cite{hamilton2017inductive}, where $\bm{\hat{A}}= \bm{A} + \bm{I}_N$ is the adjacency matrix with inserted self-loops and $\bm{\hat{D}}$ is its corresponding degree diagonal matrix with $\bm{\hat{D}_{ii}} =\sum_j \bm{\hat{A}}_{ij}$.  Our encoder can be written as:
\begin{equation}
\bm{H}_1 \!=\! \sigma(MP_1(\bm{X},\bm{A})) \quad
\bm{H}_L \!=\mathcal{E}(\bm{X,A}) \! =\!\sigma(MP_L( \bm{H}_{L-1} + \dots + \bm{H}_{1} + \bm{XWA}))
\end{equation}
where $\bm{W}$ is a learnable projection matrix and $\sigma$ is sigmoid function. 

\subsection{Classifier: Pooling and Readout Layer}
\label{pool}

To aggregate the information of each node for the graph level classification, we use Dense hierarchical pooling (DHP \cite{ying2018hierarchical}) to cluster nodes together.  After each DHP,  the number of nodes in the graph decreases. At the last level $L$, the pooling layer  is performed by a filtering matrix $\bm{\digamma} \in \mathbb{R}^{N \times Q}$. 
\begin{equation}
\bm{H}_p = \bm{\digamma}^T \bm{H}_L \qquad \bm{A}_p =  \bm{\digamma}^{T}\bm{A}\bm{\digamma} \qquad \vec{r}=\mathcal{R}(\bm{H}_p,\bm{A}_p) = \frac{1}{Q}\sum_i^Q \vec{h}_i
\end{equation} 
 produces pooled nodes $\bm{H}_p \in \mathbb{R}^{Q\times F}$ and adjacency matrix $\bm{A}_p \in \mathbb{R}^{Q\times Q}$, which generate readout vector $\vec{r}$. The final number of nodes $Q$ is predefined.  $ \bm{\digamma}$ was learned by another GraphSAGE convolutional layer optimized by a regularization loss $L_{reg} = \|\bm{A}_p, \bm{\digamma} \bm{\digamma}^{T }\|_F$, where $\| \cdot \|_F$ denotes the Frobenius norm. Readout vector $\vec{r}$ will be submitted to a MLP for obtaining final classification outputs $p$, the probability of being an ASD subject.

\subsection{Discriminator: Encouraging Good Representation}
\label{discriminator}
Following the intuition in Deep Graph Infomax \cite{velivckovic2018deep}, the good representation may not benefit from encoding counter information. In order to obtain a representation more suitable for classification, we maximize the average MI between the high-level representation and local aggregated embedding of each node, 
which favours encoding aspects of the data that are shared across the nodes and reduces noisy encoding \cite{hjelm2018learning}. The graph-level summary vector can be $\vec{s}=\sigma(\vec{r})$ as the input of discriminator, here $\sigma$ is the logistic sigmoid nonlinearity. A discriminator $\mathcal{D}(\vec{h}_i, \vec{s}): \mathbb{R}^F \times \mathbb{R}^F \rightarrow \mathbb{R}$ is used as a proxy for maximizing the MI representing the probability scores assigned to the local-global pairs. We randomly sample an instance from the opposite class as the negative sample $(\bm{X'}, \bm{A'})$. The discriminator scores summary-node representation pairs by applying a simple bi-linear scoring function \cite{velivckovic2018deep}
\begin{equation}
\mathcal{D}(\vec{h}_i,\vec{s}) = \sigma(\vec{h}_i^T\bm{M}\vec{s})
\end{equation}
where $\bm{M}$ is a learnable scoring matrix and $\sigma$ is the logistic sigmoid nonlinearity, used to convert scores into probabilities of $(\vec{h}_i, \vec{s})$ being positive.
\subsection{Loss function}
In order to learn useful, predictive representations, the Infomax loss function $L_2$ encourages nodes of the same graph to have similar representations, while enforcing that the representations of disparate nodes are highly distinct. In order to insure the performance of downstream classification, we use binary cross-entropy as the classification loss $L_1$. Therefore, the loss function of our model is written as: 
\begin{equation}
    \mathcal{L} = \underbrace{-\frac{1}{N}\sum^N_{i=1}(y_i\log(p_i)+(1-y_i)\log(1-p_i))}_{L_1} +\underbrace{ \frac{1}{2N}(\sum^{N}_{i=1}\mathbb{E}_{(\bm{X,A})}[\log \mathcal{D}(\vec{h}_i,\vec{s})]+\mathbb{E}_{(\bm{X',A'})}[1 - \log \mathcal{D}(\vec{h}'_i,\vec{s})])}_{L_2}
\end{equation}

\section{EXPERIMENT AND RESULTS}
\subsection{Data Acquisition and Preprocessing}
We tested our method on a group of 75 ASD children and 43 age and IQ-matched healthy controls collected at Yale Child Study Center \cite{li2019graph} under the "biopoint" task \cite{Kaiser07122010}. The fMRI data was preprocessed following the pipeline in Yang et al.\cite{yang2016brain}. The graph data was augmented as described in our previous work \cite{li2019graph}, resulting in 750 ASD graphs and 860 HC graphs. 
We split the data into 5 folds based on subjects. Four folds were used as training data and the left out fold was used for testing. Based on the definition in Section \ref{define}, each node attribute $\vec{v}_i \in \mathbb{R}^{10}$.  Specifically, the GLM parameters of the "biopoint task" are: $\beta_1$, coefficient of biological motion matrix; $\beta_3$, coefficient of scrambled motion matrix; $\beta_2$ and $\beta_4$,  coefficients of the previous two matrices' derivatives. 
\subsection{Experiment and Results}
We tested classifier performance on the Destrieux atlas \cite{destrieux2010automatic} (148 ROIs) using the proposed GNN with $L_1$ and $\mathcal{L}$ separately, to examine the advantage of including graph infomax loss $L_2$. In our GNN setting, $D=10$ and pooling ratios $r = 0.5$. We used the Adam optimizer with initial learning 0.001, then decreased it by a factor of $2$ every 20 epochs. We trained the network 100 epochs for all of the splits and measured the instance classification by  F-score (Table \ref{tb}). We changed the architectures by tuning either two graph convolutional layers with kernel size $(F,F)$ or one graph convolutional layer with kernel size $(F)$. $F$ was tested at 8 and 16. The regularization parameters are adjusted correspondingly to get the best performance. 

For notation convenience, we use $(\cdot)$ model and $\mathcal{L}/L_1$ model to represent the model of certain GNN architecture and corresponding training loss. Under model $(8,8)$, we could not find obvious advantage of using $\mathcal{L}$. However, if we increase the encoders' complexity to $(16,16)$, the $L_1$ model became easily overfitted while $\mathcal{L}$ model kept similar performance. This may indicate $L_2$ can perform as regularization and restrain embedding from data noise. In $(16)$ case, the $L_1$ model was underfitted, while the $\mathcal{L}$ model performed slightly better. It's probably because $L_2$ encourages encoding common nodal signals over the graph hence ignoring data noise or just because $\mathcal{L}$ model had more trainable parameters. 

\begin{figure} [ht]
   \begin{center}
   \begin{tabular}{c} 
   \includegraphics[height=3.5cm]{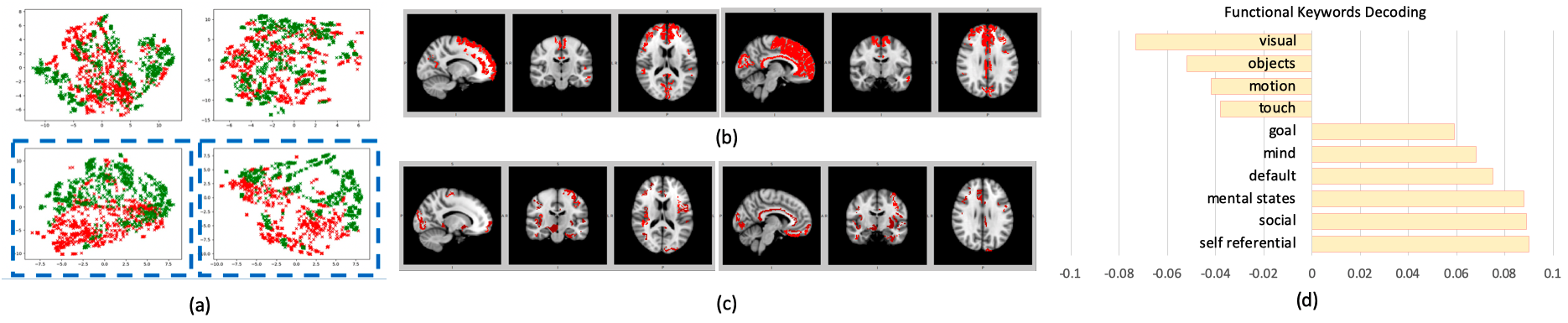}
   \end{tabular}
   \end{center}
   \caption[example] 
   { \label{results} 
Analysis of functional differences between ASD and HC. (a) shows the embedded representations of 4 brain regions visualized by t-SNE. HC is colored in green and ASD is colored in red. The top 2 regions are not separable, while the bottom two region representations are separable. (b) shows two views of the separable regions detected by our methods. (c) is the z stats of two groups by GLM. (d) shows the functional keyword decoding results of the regions in (b). }
   \end{figure}

After training, we extracted the nodal embedded vectors after the last Graph Convolutional Layer and used t-SNE \cite{maaten2008visualizing} to visualize the node presentations in 2D space. \textbf{Only with $\mathcal{L}$} did we find linearly separable nodal representations of ASD and HC for certain regions. We visually examined all the nodal representation embeddings by $L_1$ and verified they cannot be linearly separated into the two groups. We marked the regions whose Silhouette score \cite{rousseeuw1987silhouettes} was greater than 0.1 (resulting in 31 regions using $\mathcal{L}(8,8)$, shown in Fig. \ref{results} (b)) as the brain ROIs with functional difference between ASD and HC. We compared the results with GLM z-stats analysis using FSL \cite{fsl2012} (shown in Fig. \ref{results} (c)). Our proposed method marked obvious prefrontal cortex, while GLM method did not highlight those regions. Both our method and GLM analysis highlighted cingulate cortex. These regions were indicated as ASD biomarkers in Yang et al.\cite{yang2016brain} and Kaiser et al.\cite{Kaiser07122010}. Also, we used Neurosynth  \cite{yarkoni2011large} to decode the functional keywords associated with separable regions found by our methods, as shown in Fig. \ref{results} (d). The decoded functional keywords of our detected regions showed evidence that these regions might have social-, mental-, visual-related and default mind functional differences between ASD and HC group. Potentially, our proposed method can be used as a tool to identify new brain biomarkers for better understanding the underlying roots of ASD.

\begin{table}[t]	
	\centering
	\caption{Performance of different loss functions and GNN architectures (mean$\pm$ std)}
	\scalebox{0.8}{
	\begin{tabular}{p{3.5cm}|c|c|c|c|c|c}
		\hline
		\textbf{Loss + (conv-layer)}&$L_1$(16,16)&$L_1$(8,8)&$L_1$(16)&$\mathcal{L}$(16,16)&$\mathcal{L}$(8,8)&$\mathcal{L}$(16)\\ \hline
		\textbf{F-score}&$0.57\!\pm\!0.11$&$0.70\!\pm\!0.06$&$0.63\!\pm\!0.01$&$0.68\!\pm\!0.08$&$0.69\!\pm\!0.05$&$0.66\!\pm\!0.03$\\
		\hline 
	\end{tabular}
	\label{tb}	
}
\end{table}

\section{CONCLUSION}
We applied GNN to identify ASD and designed a loss function to encourage better node representation and detect separable brain regions of ASD and HC. By incorporating mutual information of local and global representations, the proposed loss function improved classification performance in certain cases. The added $L_2$ Infomax loss potentially regularizes the embedding of noisy fMRI and increases model robustness. By examining the embedded node representations, we found that ASD and HC had separable representations in regions related to default mode, social function, emotion regulation and visual function, etc. The finding is consistent with prior literature \cite{li2019graph,Kaiser07122010} and our approach could potentially discover new functional differences between ASD and HC. Overall, the proposed method provides an efficient and objective way of embedding ASD and HC brain graphs.

\label{sec:misc}

 

\bibliography{report} 
\bibliographystyle{spiebib} 

\end{document}